\documentclass{eusflat2021}

\title{\bf CG2A:
Conceptual Graphs Generation Algorithm}
\author{Adam Faci$^{a,b}$ \and Marie-Jeanne Lesot$^a$ \and Claire Laudy$^b$\\
$^a$Sorbonne Université, CNRS, LIP6, F-75005 Paris, France \email{{adam.faci,marie-jeanne.lesot}@lip6.fr} \\
$^b$LRASC, Thales, 91477 Palaiseau, France \email{claire.laudy@thalesgroup.com}}

\begin{document}

\maketitle

\begin{abstract}
  Conceptual Graphs (CGs) are a formalism to represent
  knowledge. However producing a CG database is complex. To the best
  of our knowledge, existing methods do not fully use the expressivity
  of CGs.  It is particularly troublesome as it is necessary to have
  CG databases to test and validate algorithms running on CGs.  This
  paper proposes CG2A, an algorithm to build synthetic CGs exploiting
  most of their expressivity. CG2A takes as input constraints that
  constitute ontological knowledge including a vocabulary and a set of
  CGs with some label variables, called $\gamma$-CGs, as components of
  the generated CGs. Extensions also enable the automatic generation
  of the set of $\gamma$-CGs and vocabulary to ease the database
  generation and increase variability.
  
\end{abstract}
    
\section{Introduction}

Conceptual graphs (CGs)~\cite{chein_conceptual_2008} refer to a family
of formalisms of graph-based knowledge representation, close to
existing semantic web languages such as
RDF(S)~\cite{manola2004rdf,brickley2014rdf} and
OWL~\cite{mcguinness2004owl}.  Their advantages include their data
modeling capacities, grounded on first-order logic (FOL) semantics, as
well as the possibility to manage knowledge through graph-based
operations. They differ from other graph-based semantic knowledge
representations by the clear distinction between ontological knowledge
and factual knowledge which ensures conformity of reasoning with FOL
formulas.  CGs have many applications in research and industry, eg. in
security~\cite{fu2017privacy}, semi-structure data
modeling~\cite{varga2018conceptual}, software
development~\cite{vlasenko19}, clustering~\cite{perez2019review},
music~\cite{fowler2019john} or
decision-making~\cite{tremblay2017cognitive} to name a few.  One drawback is the
major difficulty in designing a CG database without prior
expertise. It may be one of the reasons why there has been for a long
time the need of CG datasets of
quality~\cite{baget2006towards,croitoru2007conceptual}, in particular
for benchmarking. Indeed, existing CGs are either private
properties, small examples to
illustrate the formalism or specific use cases that only represent a
reduced part of the formalism, for instance with no ontological part.
As will be discussed in Section~\ref{sec:edla},
$T_{nat}$~\cite{baget_translation_2010} is a translation algorithm
from RDF(S)/OWL~\cite{manola2004rdf,brickley2014rdf,mcguinness2004owl} 
to CGs. The quality of the resulting CG base depends on the quality of
the RDF(S)/OWL input base and only a part of the CG formalism is taken
into account. The notion of quality is here mainly understood as
the combination of two criteria: 
variability, i.e. the fact that many datasets varying on several
characteristics can be generated from the same input, and
predictability, i.e. the fact that the characteristics of the
resulting database can be derived from the input without mining of the
base. Two additionnal criteria are used: expressiveness, i.e. how much of
the CG formalism is represented, and computational efficiency.

This paper proposes CG2A (Conceptual Graphs Generation Algorithm), an
algorithm generating a CG database from a set of constraints
corresponding to ontological knowledge. In essence, factual knowledge
is generated from the input ontological knowledge defined as a
vocabulary and a set of CGs with some label variables, called
$\gamma$-CGs.
The ontological knowledge thus constitutes an underlying model of the
generated dataset. A benefit is that the user has explicit
knowledge on datasets generated from this model, without
analysis of the generated datasets. It is inspired by the benchmark
generation process in the clustering community, where synthetic datasets are
generated from given data distributions that determine expected
results for a clustering algorithm running on these datasets. It has been used to validate cgSpan \cite{faci_cgSpan_21}, an algorithm proposed to mine frequent patterns in CGs.

In order to generate realistic datasets, without a total randomization
of labels and structure, ontological knowledge is
required as input. This corresponds to constraints on the generated CGs domain. Still
this input can be generated automatically from a reduced set of
numerical parameters based on three proposed extensions to the
algorithm, respectively automating the generation of the vocabulary,
the $\gamma$-CGs and the $\gamma$-CGs variables. The generated CGs
domain is therefore extended to all CGs that can be defined over the
ontological knowledge generated from the given set of numerical
parameters.
The use of these extensions requires further analysis to establish the
same quantity of ontological knowledge and thereby of expected
results, i.e. reduces predictability.
Consequently the CG2A version to use depends on the use case: on
one hand it is possible to define all input ontological knowledge or
reuse an existing one to represent a specific situation; on the other
hand the use of automatically generated input enables a swift CG
generation and leads to more variability.

Section~\ref{sec:edla} presents a short reminder about the Conceptual Graphs formalism, including the proposed $\gamma$-CGs, and a state of the art on CG datasets generation. Section~\ref{sec:CG2A} presents the proposed Conceptual Graphs Generation Algorithm as well as its randomzsation modules. Section~\ref{sec:expe} describes the conducted experimental study, detailing the proposed criteria, to measure numerically variability and efficiency, and to assess qualitatively immediate predictability and representativity of CGs formalism. Section~\ref{sec:concl} concludes the paper and discusses some directions for future works.

\section{State of the art}\label{sec:edla}
\subsection{Conceptual Graphs}\label{ssec:CG}
{Conceptual graphs~\cite{chein_conceptual_2008} are a family of formalisms for knowledge representation, made of ontological and factual knowledge. A CG is a bipartite graph representing factual knowledge referring to a vocabulary that represents the ontological knowledge.}

A vocabulary is a {5-tuple\textit{ $\mathcal{V}$ = $(T_C,T_R,\sigma,I,\tau)$}}.~$T_C$ and $T_R$, that respectively correspond to concept and relation types, are two partially ordered disjoint finite sets, where ordering corresponds to generalisation. $T_C$ contains a greatest element $\top$. Each relation type has an associated arity; which subdivides $T_R$ in subsets regrouping types of equal arity. $\sigma$ is a mapping associating a signature with each relation type, i.e. a function with constraints on the type of arguments, where a more specific relation type is mapped with a more restrictive signature respectively for each argument. $\sigma(r)$ returns $(t_1, \ldots, t_n)$ where $n$ is $r$ arity and the $t_i$ are elements of $T_C$.
For $c$ connected to $r$, $\sigma(r)(c)$ denotes the type restriction matching $c$. $I$ is a set of individual markers used to instantiate concept nodes.~$\tau$ is a mapping from $I$ to $T_C$ that defines the type instantiated by each individual marker.

 A CG is a bipartite labeled multigraph represented as a 4-tuple
 \textit{G = (C,R,E,label)} defined over such a
 vocabulary~$\mathcal{V}$.  $C$ and $R$ correspond to concept and
 relation nodes, $E$ denotes the set of the edges connecting elements
 of $C$ and~$R$. $label$ is a labelling function from $C$ to
 $T_C\times I$ and from $R$ to $T_R$. For any $r\in R$,
 $label(r)=t_r\in T_R$ is the type of $r$ and for any $c\in C$,
 $label(c)=(t_c,i_c)\in T_C\times I$ where $t_c$ is the type of $c$
 and $i_c$ is the optional individual marker of $c$.

 We extend this formalism to represent CGs where some labels are
 replaced with variables, named $\gamma$-Conceptual Graphs and
 inspired by $\lambda$-BGs from the CG
 formalism~\cite{chein_conceptual_2008}.~A $\gamma$-CG
 $\Gamma=((v_1,D_1) \ldots (v_n,D_n)) G$, $n \geq 1$ is a conceptual
 graph $G$ with $n$ variables $v_i$ and their respective domains
 $D_i$. Each variable $v_i$ is assigned to a label of~$G$, either a
 relation type label, concept type label or marker label.  It is
 illustrated in Fig.~\ref{fig:CG2AAutoVar} where $v_1$, $v_2$ and
 $v_3$ are respectively assigned to a concept type, marker and
 relation type.  For a variable $v_i$ associated with a relation type
 of $r$ with $label(r)=t_r$, its domain~$D_i$ is a subset of $T_R$
 reduced to relation types of same arity, i.e.
 $D_i = \{t \in~T_R, arity(t) = arity(t_r)\}$.  For a variable $v_i$
 associated with a concept type of $c$ with $label(c)=(t_c,i_c)$, its
 domain $D_i$ is a subset of $T_C$ reduced to the concept types
 respecting all constraints imposed by the signatures of connected
 relation nodes, i.e.
 $D_i = \{t \in~T_C, \forall~r \in~R, (c,r) \in~E, t \leq~
 \sigma(r)(c)\}$.  For a variable $v_i$ associated with a marker
 $m_i$, the domain is a subset of $I$ reduced to markers of same or
 more specific concept types, i.e.
 $D_i = \{m \in~I, type(m) \leq type(m_i)\}$.

Finally we define a neighborhood as a node and its connected nodes, for instance a relation node and its connected concept nodes.

\subsection{Conceptual graphs data generation}

To the best of our knowledge, there is no CG dataset of quality
available. Available CG datasets are for instance based on
flat hierarchy, i.e. with no order defined between types, as the
conceptual graph data-set for
NLP/NLU~\cite{elseidy2014grami}\footnote{\url{https://github.com/alexge233/conceptual_graph_set}},
or even actually not consistent with the CG formalism. There exit CG
datasets in industry but they remain company property, as they may be
the result of intense work 
and may contain private data.

CG datasets can be obtained as the result of 
translation algorithms to generate them from a dataset respecting
another formalism. The main differences with a proper generation
algorithm are that the goal is different and that the resulting dataset
depends on the chosen input dataset and its formalism.  $T_{3}$
and $T_{nat}$~\cite{baget_translation_2010} are existing algorithms
translating knowledge datasets expressed in the
RDF(S)/OWL~\cite{manola2004rdf,brickley2014rdf,mcguinness2004owl}
formalism to knowledge datasets expressed in the CG formalism. They
are implemented in CoGUi\footnote{\url{http://www.lirmm.fr/cogui/}}, a
tool to visualize and manipulate CGs.  Their main validation criterion
is the equivalence between reasoning in RDF(S)/OWL before translation
and reasoning in the CGs formalism after translation: they aim at
ensuring that the same conclusions are deduced from the same premises
in both datasets, and that reasoning remains identical when translating
back to RDF(S)/OWL. In this regard, $T_{3}$ is a sound and complete
translation w.r.t. RDF(S) but not intuitive visually. Indeed, it represents the RDF(S)/OWL triplets
constituted of subject, object and predicate by a blank relation node
linking these three elements as concept nodes. As a consequence the
fact that relations in CGs correspond to relations between entities is
not represented. It is more intuitive to represent relations nodes
connecting concepts nodes as predicates linking subject and
object, as is the case of~$T_{nat}$.

In addition $T_{nat}$ focuses on exploiting the separation between background knowledge and factual knowledge by translating the predicates as binary relation nodes linking the subject and object, both translated as concept nodes. This translation ensures two properties that enable a better representation of CGs but hinders the reasoning equivalence. First, a \textit{separability condition} has to be satisfied by the input RDF(S)/OWL dataset: it states that any entity in the knowledge base appears either as a class, a property or an instance (in the RDF(S) sense). Otherwise the entity is considered ambiguous and different choices are made depending on the situation: if a violation of this separation requirement between classes and properties occurs, the {ambiguous predicates are ignored}: if a violation occurs between {classes and instances, or properties and instances}, the triples involving the ambiguous entity as an instance are ignored.
Second, a distinction between ontological and factual triples is performed to populate either the vocabulary or the conceptual graphs when a new triple is processed. This distinction stems from {the flexibility of RDF(S) that does not impose a clear distinction} between factual and ontological knowledge.
A particularity of GC databases constituted with~$T_{nat}$ is that only relations of arity~2 are built
because of RDF(S) restrictions. This drawback is minimised by the fact that relation of arity greater than 2 can always be brought back to a set of relations of arity 2, and conversely. This is immediate considering that CGs are graph-based representations of first-order logic formulas and that relations correspond to atomic formulas, which are 2-decomposable~\cite{jeavons1998constraints}.

\section{CG2A: generation from a set of constraints}\label{sec:CG2A}

CG2A is a three step algorithm that generates a CG dataset from
ontological knowledge. It ensures representativity of the CG formalism
as well as variability and immediate predictability of the generated
base characteristics. First CG2A generates a CG by randomly combining
input $\gamma$-CGs until reaching a specified minimum size. Then
variables are assigned random values from their respective
domains. Finally the nodes in the generated CGs with the same
individual marker are merged to increase the connectivity of the resulting CG. CG2A iterates until a specified number of CGs is reached.
This section first describes CG2A input and details its three steps. It then presents its extension modules automating the generation of input.

\subsection{Input}\label{ssec:input}

CG2A, in its default mode, takes five parameters. They include the number of CGs to be generated, \textit{maxCGs}, the minimum size, in number of nodes, for each generated CG, \textit{minSize}, and the maximum number of specializations to be operated on each variable assigned to a type label, \textit{maxSpe}. They are used in the stopping conditions of the algorithm. The two other parameters are a vocabulary $\mathcal{V}$ and a set of $\gamma$-CGs $\mathcal{G}$, as detailed hereinafter.

The vocabulary, as formally presented in Section~\ref{ssec:CG}, contains a hierarchy on concept types, a hierarchy on relation types and a set of signatures corresponding to the relation types in the hierarchy. The individual markers set is populated during generation when a concept node is instantiated. The set~$\mathcal{G}$ of $\gamma$-CGs are the components of the generated CGs. Compared to a classic CG, as described in Section~\ref{ssec:CG}, some labels are replaced with a variable referring to a list of values from the vocabulary. Thus $\gamma$-CGs are configurable constraints.

\subsection{Proposed algorithm}\label{ssec:CG2A}

Fig.~\ref{fig:CG2A} gives the pseudo-code of CG2A, commented below: 
CG2A generates sets of CGs by randomly combining elements from the set~$\mathcal{G}$ of input $\gamma$-CGs into bigger CGs.
Let $G_c=(C_c,R_c,E_c,label_c)$ be the currently generated CG and
$\Gamma = ((v_1,D_1) \ldots (v_n,D_n)) G=(C,R,E,label)$ be a
$\gamma$-CG from the input set $\mathcal{G}$. First $G$ variables are
instantiated with values from their domains, and the ones assigned to
type labels are specialized from 0 to \textit{maxSpe} times using
hierarchies from~$\mathcal{V}$. Then $G_o = (C_c\uplus C, R_c\cup R,
E_c\uplus E, label_c\uplus label)$ is formed from the join of $G_c$
and $G$ where $\uplus$, based on coreferent nodes
merge~\cite{chein2004concept}, is a specific union whose differences
follow: if there are elements of $C_c$ with an individual marker
similar to one of $C$, only the most specialised is kept. Then
neighborhoods are merged so that both neighborhoods are connected to
the resulting node, i.e. elements of $E_c$ and $E$ corresponding to
the two merged nodes are reassigned to the resulting node.

\begin{figure}
    Input: $\mathcal{V}$ = $(T_C,T_R,\sigma,I,\tau)$, $\mathcal{G}$,
    \\\phantom{Input:} \textit{maxCGs}, \textit{minSize}, \textit{maxSpe}.
    \begin{itemize}
        \setlength\itemsep{0.5em}
        \item Initialize $\mathcal{G}_o$ to an empty set
        \item Iterate until size($\mathcal{G}_o$)$\geq$\textit{maxCGs}
        \begin{itemize}
            \setlength\itemsep{0.2em}
            \item Initialize $G_c=(C_c,R_c,E_c,label_c)$ to an empty CG
            \item Iterate until $size(G_c)\geq minSize$
            \begin{enumerate}
                \item Get $(v_1,\ldots,v_n)G=(C,R,E,label)$ in $\mathcal{G}$
                \item Attribute value to each variable $v_i$
                \item Specialize each type label var from 0 to \textit{maxSpe} times
                \item $G_c = Join(G_c,G)$
            \end{enumerate}
            \item Add $G_c$ to $\mathcal{G}_o$
        \end{itemize}
        \item Return $\mathcal{G}_o$
    \end{itemize}
    \caption{Pseudo-code of the proposed CG2A.}
    \label{fig:CG2A}
\end{figure}

\begin{figure}[t]
\begin{center}
\includegraphics[width=0.9\columnwidth]{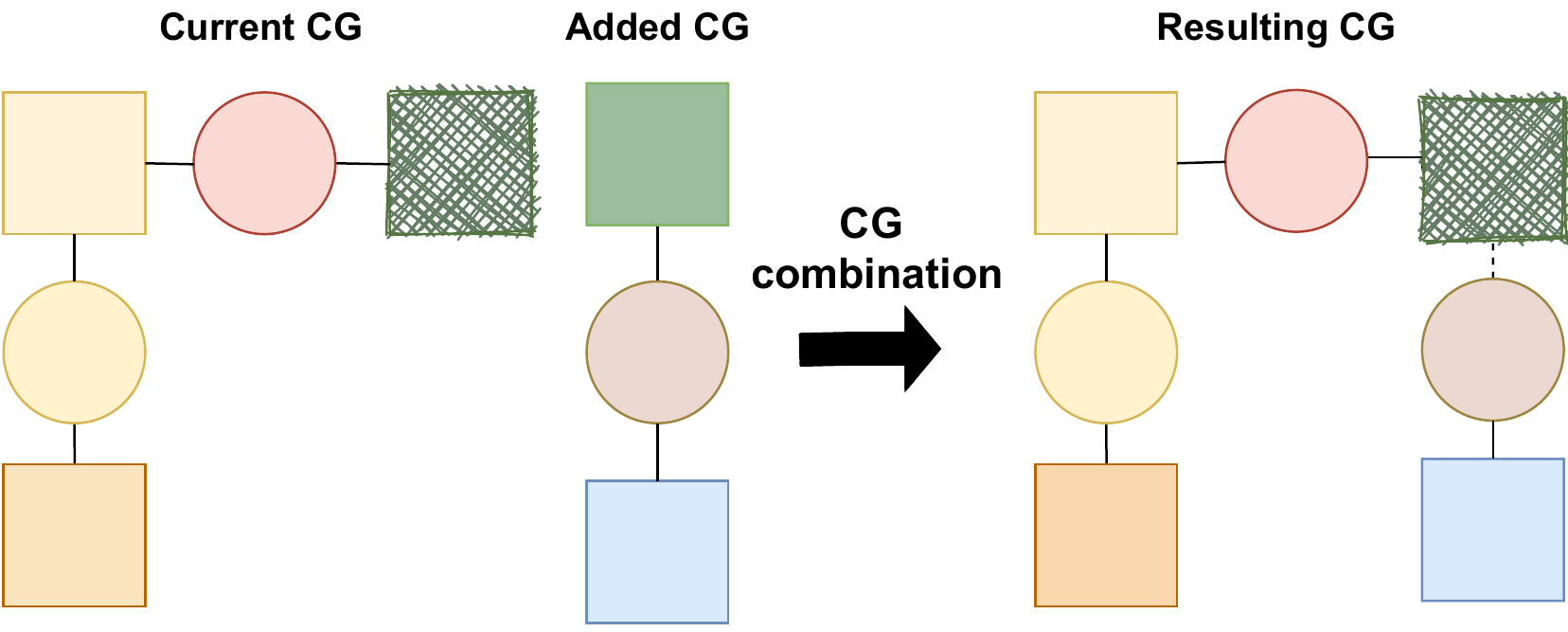}
  \caption{ CG join step in CG2A. In this representation, concept nodes are squares and relation nodes are circles.}
\label{fig:CG2AFusion}
\end{center}
\end{figure}

This join operator is illustrated in Fig.~\ref{fig:CG2AFusion}, where
the node colour indicates their associated markers: the two green
nodes, resp. at the right end of the current CG and at the top of the
added CG, are merged. They are not necessarily of same type; the most
specific type is retained, indeed as illustrated in the example, the 
connected signatures enforce a specialisation of this
type.

Without operator $\uplus$, the algorithm would obtain for each
generated CG a set of unconnected instantiated elements from
$\mathcal{G}$. The connectivity of the resulting CGs thus depends on
the number of common nodes. There are other techniques available for
graph fusion based on the join
operator~\cite{laudy2007high,chein2014conceptual}, but this simple
fusion operator based on coreferent nodes merge operator is sufficient
in this case.

CG2A stops CGs combinations upon reaching the desired minimum size,
\textit{minSize} and stops generation upon reaching the desired number
of generated CGs, \textit{maxCGs}. Since CGs of potentially several
nodes are added at the same time, the resulting CGs are typically
greater than \textit{minSize}.

The advantages of using $\gamma$-CGs instead of directly defining many
variants of a CG is that the process is automatic and that from one
designed $\gamma$-CG, many can be generated while keeping its
structure and its semantic. As a consequence CG2A guarantees
variability from one input as well as predictability thanks to
knowledge of input $\gamma$-CGs and their characteristics.

\subsection{
  Input generation to increase variability}\label{ssec:random}

This section presents three modules to generate automatically the
input so as to increase variability and ease the generation.  All
mentioned numerical parameters   can be replaced by mean
and standard deviation, and drawn from  
the associated normal distribution. 

\subsubsection{Automatic generation of vocabulary }\label{sssec:randVoc}
This module generates automatically the vocabulary~$\mathcal V$ from four
parameters: the desired depth of hierarchies both for concept and
relation types, the maximum number of children of each node of the
hierarchy and the number of individual markers for each concept
type. This generation is random, however the four parameters ensure a
number of fixed characteristics in the resulting vocabulary.

\begin{figure}[t]
\begin{center}
\includegraphics[width=\columnwidth]{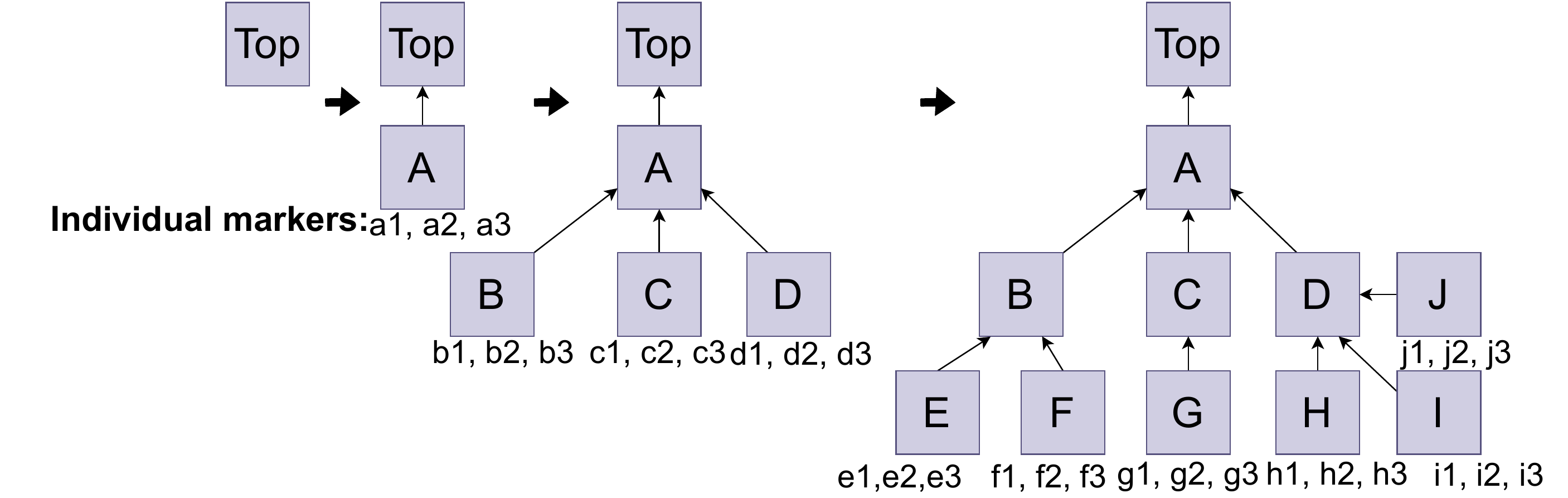}
  \caption{ Automatic generation of a hierarchy of concept types, here performed in three steps.
  Parameters are: Depth = 4; Maximum number of children = 3; Number of individual markers per type = 3.}
  \label{fig:CG2AAutoVoc}
\end{center}
\end{figure}

\begin{figure}[t]
\begin{center}
\includegraphics[width=\columnwidth]{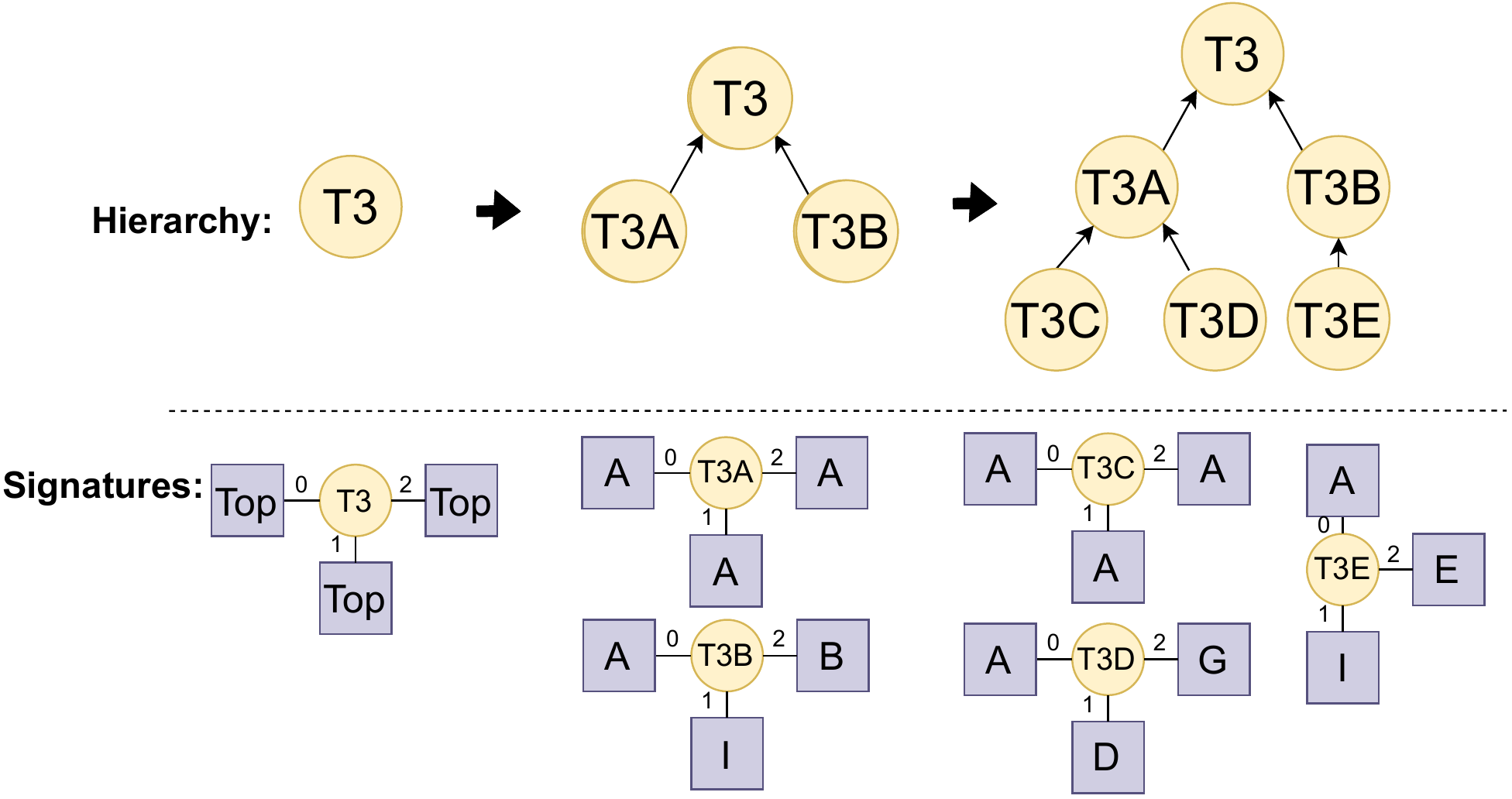}
  \caption{ Automatic generation of a hierarchy of relation types with their signatures, here performed in two steps.
  Parameters are: Depth = 3; Maximum number of children = 3.}
  \label{fig:CG2AAutoSign}
\end{center}
\end{figure}

As illustrated on Fig.~\ref{fig:CG2AAutoVoc} and
\ref{fig:CG2AAutoSign}, for concepts and relations respectively, a
hierarchical structure is generated until the desired depth is reached
and random unique labels are assigned to each node of the hierarchy.
The hierarchy of concept types is a rooted tree with the most general
type $\top$, denoted "Top" in Fig.~\ref{fig:CG2AAutoVoc}. For each
concept type, a list of individual markers is generated.  For relation
types, a top type is respectively defined for each arity, e.g. denoted
$T_3$ for the case of arity 3 illustrated in
Fig.~\ref{fig:CG2AAutoSign}.  Then signatures are defined using the
previously generated hierarchy of concepts for each relation type,
with each relation top type having a default signature with only
$\top$ as concept type restriction. A more specific relation type has
a more restrictive signature, meaning that the specified restrictions
require an identical or more specific concept type. It is illustrated
in Fig.~\ref{fig:CG2AAutoSign} where at each step the hierarchy is
deepened and signatures are defined as identical or more restrictive
than signatures of more general relation types.

\subsubsection{Automatic generation of input $\gamma$-CGs}\label{sssec:randCG}
\begin{figure}[t]
\begin{center}
\includegraphics[width=\columnwidth]{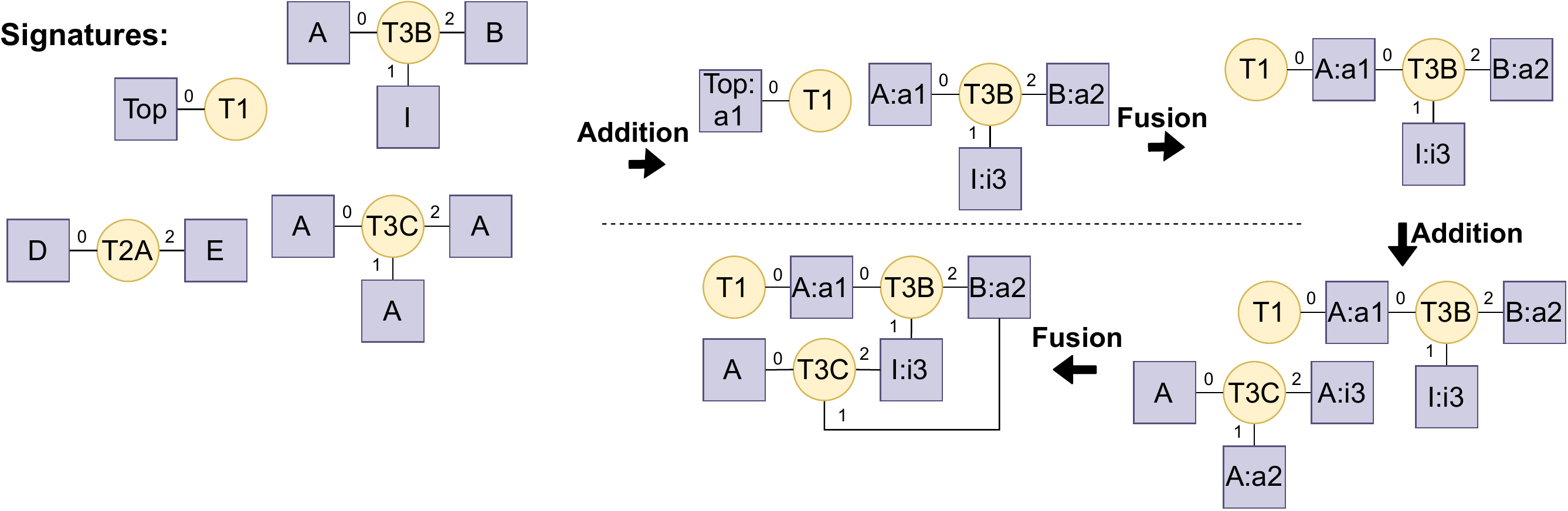}
  \caption{Automatic generation of input $\gamma$-CGs. }
  \label{fig:CG2AAutoCG}
\end{center}
\end{figure}

This module generates automatically a set of input $\gamma$-CGs so as to define $\mathcal{G}$, as illustrated in Fig.~\ref{fig:CG2AAutoCG}. The generated $\gamma$-CGs actually have no defined variable, but as this module can be used independently, one can subsequently define variables manually or use automatic generation. This module takes as input a vocabulary~$\mathcal{V}$ (possibly generated automatically using the module described in the previous subsection~\ref{sssec:randVoc}) and two numbers: the number of $\gamma$-CGs to be generated and their minimum size. While designing CGs requires proficiency in the formalism and also requires to respect of the vocabulary constraints, only two numbers restrict the domain of $\gamma$-CGs generation.

The $\gamma$-CGs module is similar to applying CG2A to the set of signatures taken from the input vocabulary $\mathcal{V}$. The only difference is that each label is treated as a variable with no constraint, so that they all have a randomly attributed label and are randomly specialized.
    
\subsubsection{Automatic generation of variables}\label{sssec:randVar}

This module generates variables in input $\gamma$-CGs. Instead of having to choose which labels are variables and their respective domains, this module takes a set of $\gamma$-CGs and a matching vocabulary as input, which can be generated with the previous modules, as well as five numbers: the numbers of concept types, relation types and individual marker variables per CG, the number of values per variable and the number of specialisations. It may be run even if variables have already been defined in the $\gamma$-CGs to increase their number.

\begin{figure}[t]
\begin{center}
\includegraphics[width=\columnwidth]{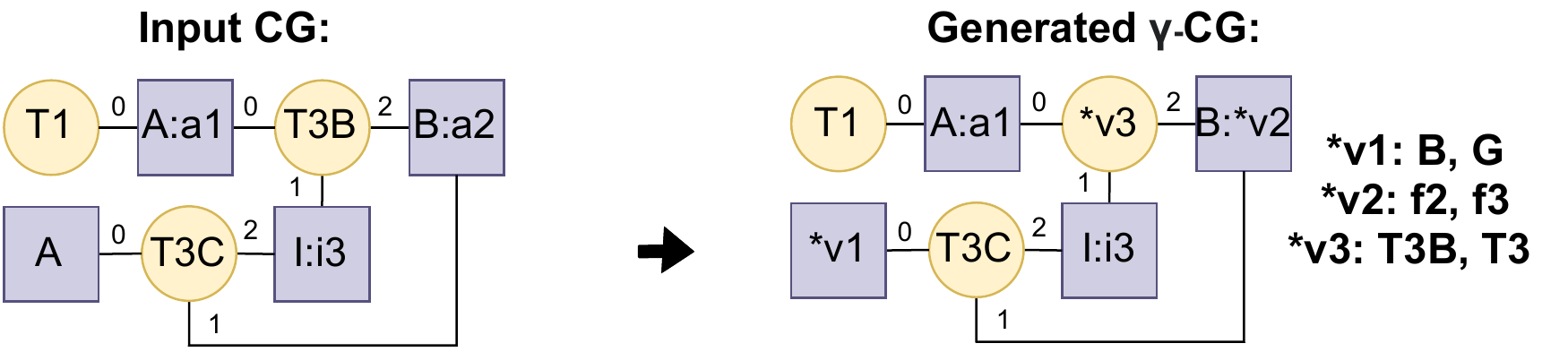}
  \caption{ Automatic generation of $\gamma$-CGs variables.}
  \label{fig:CG2AAutoVar}
\end{center}
\end{figure}

First, for each CG, variables are attributed to a relation type, a
concept type or an individual marker. Then a list of values is
associated with each variable. Fig.~\ref{fig:CG2AAutoVar} illustrates
this operation with the variables $v_1,v_2$ and $v_3$.  For a relation
type as $v_3$ in Fig.~\ref{fig:CG2AAutoVar}, the module chooses from
relations with the same arity and identical or less restrictive
concept types.  For a concept type as $v_1$ in
Fig.~\ref{fig:CG2AAutoVar}, the module chooses from concept types
equal to or more specific than the ones compatible with the signatures of
the neighborhood.  For an individual marker as $v_2$ in
Fig.~\ref{fig:CG2AAutoVar}, it chooses from individual markers with an
assigned concept type equal to or more specific than the concept node
type.  Because of these restrictions, first relation type variables
are defined, then concept type variables and finally individual marker
variables.  Then, all assigned variables corresponding to type labels
are specialized a number of times up to the number of specialisations
parameter.

\section{Experimental study}\label{sec:expe}

This section presents the experimental study conducted to evaluate CG2A both per se and compared to existing translation techniques. 
The input used in these experiments as well as the desired properties of variability, predictability, representativity and efficiency are presented in the following subsection. Then results are subsequently examined in view of each criterion.

\subsection{Experimental protocol}\label{ssec:prot}

We consider four criteria to optimize for 
a data generation algorithm.  First the algorithm has to enable
\emph{variability} in the generated data from one input. Indeed from a
unique ensemble of ontological knowledge the possibility to produce
datasets with various sizes and characteristics may be required to
assess the breadth of corresponding possible facts.  Second it has to
provide a certain level of \emph{immediate predictability}. This means that
from a given data generation algorithm, denoted dGenA, 
the expected results of a data mining algorithm, denoted dMinA,
run on a database generated by dGenA can be defined. Obviously the
expected results will differ depending on dMinA goal and whether all
that can be deduced from the dGenA input is relevant regarding this
goal. This criterion is essential to enable dMinA validation but is
difficult to quantify.  Third the generation algorithm has to achieve
\emph{representativity}, i.e. exploit as much of the CG formalism as
possible. Any fragment that cannot be represented limits the use of
the generation algorithm to the subset of situations where this fragment is
unnecessary.  Fourth the \emph{computational time} 
has to be minimised. Indeed it may be crucial to
obtain quickly a set of examples datasets when given ontological
knowledge. Besides many generations may be required while testing
parameters variations to satisfy expectations.

These criteria are used to compare CG2A with $T_{nat}$ using the
following configuration.  The input of $T_{nat}$ is an RDF(S)/OWL
dataset\footnote{\url{http://www.semanticbible.com/ntn/ntn-view.html}},
modified to resolve some issues when parsing for translation. The
modified dataset includes an ontology constituted of 39 concept types
and 35 relation types organized in a hierarchy with a depth of 5 and
between 1 and 3 children for each non-leaf type. An option proposed by
$T_{nat}$ to split CGs in several connected components has been used,
so that each resulting CG is a connected graph. Otherwise it results
in one big unconnected CG. The input for CG2A 
matches the characteristics of~$T_{nat}$ produced base to ensure that
we mostly evaluate the influence of the algorithmic part rather than
the variability induced by parameters. Running $T_3$ on the same
dataset results in a unique CG of about 6000 vertices and no ontology
(other than RDF/RDF(S) knowledge). All relation nodes are the "triple"
relation node to connect elements of a triple. As it does not lead to 
many CGs nor a proper ontology, $T_3$ is irrelevant for our concerns
and in consequence its results are not used in what follows.

The first row of Tab.~\ref{table1} displays the results of $T_{nat}$
and the following ones display the average results across 100 runs on CG2A and
its variants with each extension module individually (Auto Voc, Auto
$\gamma$CG and Auto Var respectively) and CG2A with all extension modules
(Full Auto).

\subsection{Variability results}

In order to assess numerically the notion of variability, we consider
the following criteria: the average size of generated CGs in number of
nodes, denoted~NbN, and the average number of unique labels, denoted
NbL, in one CG, both with their standard deviation, and the
distribution of relation arity from 1 to 3, denoted Ar1, Ar2 and Ar3.
These criteria are defined for the sake of comparison with existing
techniques providing CG datasets. They are not optimized or even considered by the
translation algorithm $T_{nat}$ that aims at maximizing conformity of reasoning between the input and
output databases. CG2A and its variants can represent
relation of arities greater than 3, but the use Ar1, Ar2 and Ar3 seems
to suffice for the presented experiments.

It can first be observed that the translation of $T_{nat}$ results in
a huge CG of hundreds of nodes and many CGs of a few nodes. This is
why in Tab.~\ref{table1} the standard deviation of NbN is
significantly more important than the average NbN, and the average NbL
is relatively small.  Moreover Ar1 and Ar3 are zero, which is due to
the fact that the RDF(S)/OWL languages do not represent relations of
arities different than 2.

CG2A leads to significantly smaller standard deviations, and the resulting
CGs characteristics are close to the input parameters.

It is expected that the use of automatically generated input leads to
more variability in the expected results. It can be observed that,
indeed, the use of random vocabulary increases NbL standard
deviation and relation arities while the use of random CGs increases
the standard deviation of NbN and arities. The results of the fully
automated CG2A combine these consequences.

\begin{table*}[t]
\begin{center}
\vspace{1ex}
 \begin{tabular}{|c|c|c|c|c|c|}
        \hline
        \textbf{Test} & \textbf{NbN (avg. $\pm$ sd.)} & \textbf{NbL (avg. $\pm$ sd.)} & \textbf{Ar1} & \textbf{Ar2} & \textbf{Ar3}\\
        \hline
        $T_{nat}$ & 15.2 $\pm$ 321 & 3 $\pm$ 1 & 0 & 3 & 0\\
        \hline
        CG2A & 36.3 $\pm$ 4 & 22.5 $\pm$ 4 & 0.5 & 44 & 3\\
        \hline
        Auto Voc & 33 $\pm$ 3.5 & 55 $\pm$ 14 & 4 & 34 & 9\\
        \hline
        Auto $\gamma$CG & 39.9 $\pm$ 2 & 22 $\pm$ 4.1 & 6 & 22 & 31\\
        \hline
        Auto Var & 35.3 $\pm$ 4 & 32 $\pm$ 7 & 0.4 & 42 & 7\\
        \hline
        Full Auto & 35 $\pm$ 4 & 67 $\pm$ 17 & 8 & 33 & 26\\
        \hline
    \end{tabular}
\caption{Results for one run of $T_{nat}$ and different versions of CG2A averaged across 100 runs (see experimental protocol in Sec.~\ref{ssec:prot}).}
\label{table1}
\end{center}
\end{table*}

\subsection{Predictability results}

Predictability refers to the possibility to define a priori the
results a data mining algorithm is expected to obtain when run on a
generated data set.  It can be put in balance with the cost of the
specific resources to deploy and efforts to undertake to define the
input, in particular the ontological knowledge.

$T_{nat}$ possesses the advantage that it generates a CG database from
a  RDF(S)/OWL dataset, without requiring any prior
knowledge on this dataset. However this implies that without mining
the input dataset first, $T_{nat}$ cannot be considered as
predictable. As such, it does not meet the immediate predictability
aim.

CG2A can be considered as predictable as the generated CG are defined as combination of the input $\gamma$-CG that are defined over the input vocabulary: $\mathcal G$ together with $\mathcal V$ determine the expected results whose characteristics are known.

CG2A used with automatic generation of vocabulary or $\gamma$-CGs changes the nature of the expected results, that are defined in terms of their general characteristics rather than specific information. 
As compared to CG2A, in the case of Auto~Voc, only the general characteristics of the  vocabulary are known, its specificities are not. In the case of Auto~$\gamma$-CG, only the general characteristics of the components used to build the generated CG are known (number and size of the $\gamma$-CGs, as well as the signatures they are built on): Auto~$\gamma$-CG adds to the CG2A variation from $\gamma$-CGs to generated CGs another variation,  from the signatures to the input $\gamma$-CGs. 

CG2A used with automatic variable generation, Auto~Var, does not modify predictability
significantly. We consider that automatic variables slightly reduce
the immediate predictability by increasing variability.

The fully automated CG2A variant, that includes the three
random modules, combines their respective properties and defines expected results in terms of their general characteristics. 
Overall, the results are significant when compared with $T_{nat}$:
there is much more variability with CG2A, and while CG2A has to cope
with a balance between variability and immediate predictability,
$T_{nat}$ does not enable immediate expected results.

\subsection{Representativity results}

As discussed earlier, $T_{nat}$ outputs CG databases only including relations of arity 2,  that are unbalanced with respect to ontological or factual knowledge, i.e. that mostly comprehend one of the two types. Yet, if can be argued that the advantage of disposing of relations with varying arity only is a question of perspective or reformulation and that the lack of balance is due to the considered input RDF(S)/OWL knowledge bases. Similarly, the fact that $T_{nat}$ often results in bases constituted of one or two huge conceptual graphs and the rest containing only a few nodes is mostly due to the available RDF(S)/OWL bases, rather than the algorithm itself.

CG2A and its variants more naturally avoid these drawbacks. They enable the representation of most of the CG formalism as reminded in Sec.~\ref{sec:edla}.  CG2A retains most of $T_{nat}$
advantages by using the CoGui formalism and adds the possibility to
generate a large proportion of relation nodes with various arities,
and to have both a wide vocabulary and a considerable quantity of CGs,
i.e. both ontological and factual knowledge.  Besides when defining
the input, e.g. the characteristics of the vocabulary, the user can determine the extent of  the CGs formalisms  that is exploited, which is one main advantage of CG2A. 
Generally speaking, CG2A ensures that the user can choose more precisely the characteristics of the resulting base.

\subsection{Efficiency results}

In the conducted experiments,  depending on the stopping conditions parameters, most CG2A runs last less than one second and never exceed 5 seconds. The use of the automatic generation modules increases the computational time, with a factor~2, however the time spent to design input without these modules is not accounted for. $T_{nat}$ generation lasts much longer, with factor of three up to ten depending on the size of the input base. At some point if the input is too massive, $T_{nat}$ aborts so the tests could not be pursued.

\section{Conclusion}\label{sec:concl}

This papers proposes CG2A, an algorithm to produce Conceptual Graphs.
CG2A enables more variability in the generated dataset than any other known method as it offers a lot of variance in the size and labels of CGs as well as a reasonable proportion of relation nodes with an arity different than 2 in the generated CGs. As such, numerous different situations can be tested through the use of CG2A, either a strongly constrained domain to test a specific case or a more relaxed generation to test a broad variety of situations. In addition the CG formalism is well represented with deep hierarchies, variation in the signatures and various arities of relation nodes.
Finally, when using a different method to obtain a CG dataset, it is not possible to define expected results without first mining the dataset or having prior knowledge on this dataset.

Ongoing works aim at extending CG2A to generate more complex CGs, e.g.  nested or fuzzy CGs. Another direction considers a different possibility in terms of predictability, expressed as a desired distribution over CGs parameters so as to ensure that the resulting dataset respects such a distribution.

\bibliographystyle{eusflat2021}
\bibliography{bib}

\end{document}